\documentclass[aps,pra,onecolumn,groupedaddress,superscriptaddress,tightenlines,showpacs,nofootinbib]{revtex4}
\usepackage{amssymb}
\usepackage{color,graphicx}
\usepackage{lmodern}
\usepackage{amsmath}
\usepackage{amsbsy}
\usepackage{float}
\usepackage{bbm}
\usepackage{bm}
\usepackage{epsfig}
\usepackage{braket}
\usepackage{graphics}

\newcommand{\bq}{\begin{equation}} \newcommand{\eq}{\end{equation}}
\newcommand{\bqali}{\bq\begin{aligned}}
\newcommand{\eqali}{\end{aligned}\eq}
\newcommand{\bqn}{\begin{equation*}}
\newcommand{\eqn}{\end{equation*}}
\newcommand\D{\operatorname{d}}

\newcommand\Kerr{\operatorname{Kerr}}
\newcommand\re{\operatorname{Re}} 

\newcommand\Int{\operatorname{int}}
\newcommand\tot{\operatorname{tot}}
\newcommand\cat{\operatorname{cat}}
\newcommand\BSA{\operatorname{BS1}}
\newcommand\BSB{\operatorname{BS2}}

\newcommand\ph{\operatorname{ph}}
\newcommand\DD{\operatorname{D}}
\newcommand\Tr{\operatorname{Tr}}
\newcommand\Th{\operatorname{Th}}

\begin{document}
\title{Bringing Schr\"{o}dinger's Cat to Life with Non-Equilibrium Respiration}

\author{Arash Tirandaz}
\email[Corresponding Author:~]{arash85201@ipm.ir}
\affiliation{School of Biological Sciences, Institute for Research in Fundamental Sciences (IPM), P.O.Box 19395-5531, Tehran, Iran}

\author{Hamid Reza Naeij}
\email[]{naeij@alum.sharif.edu}
\affiliation{Research Group on Foundations of Quantum Theory and Information,
Department of Chemistry, Sharif University of Technology
P.O.Box 11365-9516, Tehran, Iran}
\author{Afshin Shafiee}
\email[]{shafiee@sharif.edu}
\affiliation{Research Group on Foundations of Quantum Theory and Information,
Department of Chemistry, Sharif University of Technology
P.O.Box 11365-9516, Tehran, Iran}

\begin{abstract}

In this study, we have proposed a method based on non-equilibrium effects to generate the superposition of macroscopically distinguishable quantum states, known as Schr\"{o}dinger cat states, by using a Mach-Zehnder interferometry type experiment. Interaction of the input number state with a Kerr medium in the presence of a couple of heat baths in different temperatures in interaction picture and without imposing Markov assumption is considered. We have shown that the study of dynamics of the cat states under non-equilibrium condition open a way for the robustness of quantum features against the destructive role of the environment even at high temperature limit. It is verified that mutual influence of the environments, far from equilibrium, on the open system, makes it possible to revive quantum beats for longer time intervals. Moreover, we have probed how the traits of the environment, like its temperature and the Ohmic, super-Ohmic or sub-Ohmic functionality of the spectral density, may affect the pattern of the oscillation between alive or dead states of the cat.

\end{abstract}

\maketitle
\textbf{keywords}: Schr\"{o}dinger Cat States, Non-Equilibrium Quantum Effects, Mach-Zehnder Interferometer, Kerr Interaction.

\section{Introduction}

In recent years, there have been much attentions to the superposition of macroscopically distinguishable quantum states, known as Schr\"{o}dinger cat states (SCSs). In quantum optics, a SCS is defined as a superposition of two coherent states with opposite phase \cite{Johnson}. Such superpositions may show various non-classical properties such as sub-Poissonian statistics and squeezing due to the interference between the components \cite{Foldesi,Janszky}.

SCSs play an important role not only in the study of the conceptual foundations of quantum mechanics \cite{Leggett}, such as quantum to classical transition and quantum decoherence \cite{Zurek}, but also for various applications in quantum information processing such as quantum metrology \cite{Ralph,Munro,Gilchrist}, quantum teleportation \cite{van Enk,Jeong 1,Wang} and quantum computation \cite{Jeong 2,Ralph 2}.

Generation of SCSs, in theoretical and experimental contexts, has always been an important problem in the quantum optics researches. So far, many works have been done to generate these states.  Brune {\it et al.} used dispersive atom-field coupling to generate such states \cite{Brune}. These states have been generated with cavity quantum electrodynamics \cite{Brune 2, Davidovich}. Recently, Liao proposed a method to generate a macroscopic superposition of coherent states in a quantum harmonic oscillator coupled to a quantum bit  via a conditional displacement mechanism \cite{Liao}. Also, the generation of such states by photon subtraction is widely used due to its simplicity \cite{Asavanant}.

Moreover, using the non-linear optical processes based on the Kerr effect is a common approach to generate SCSs. In the one of the earliest works, Yurke and Stoler showed that self-modulation of coherent light interacted with a Kerr medium could lead to generate such states \cite{Yurke}. Gerry has proposed methods for generating of such states  by using a Mach-Zehnder (MZ) interferometer with a Kerr medium \cite{Gerry 1,Gerry 2}. The obtained states from this approach are called the Yurke-Stoler states. The important point about Yurke-Stoler states is that the macroscopic features of them are reinforced than the typical coherent states. So, the Yurke-Stoler states are more valid representatives of macroscopic quantum states. Moreover, these methods have a very important advantage and it is that they are independent of the conditional measurements \cite{Paris}.

However, there is an important problem in generation of SCSs. The superposition of macroscopically quantum states intercat with the environment strongly. So, these states are very sensitive to decoherence effects, and as a result their generation is very challenging \cite{Leibfried,Ourjoumtsev}.

In this regard, many studies have been proposed to generate SCSs considering the decoherence effects. Bose showed that preparation of the non-classical states in the cavities in the presence of the environment \cite{Bose}. Braun introduced an experiment in which the extraordinarily slow decoherence should be observable in generation of the superposition of macroscopically distinct quantum states \cite{Braun}. A numerical study has been done to show the decoherence effects on the macroscopic entanglement generation \cite{Jeong 3}. Yang introduced a method for generating an optical SCS in a cavity under decoherence effects \cite{Yang}.  It is noteworthy that in most of the works have been done in this field, the master equations are used to describe the decoherence effects.

In the present study, we investigate the generation of the optical SCSs using a MZ interferometer containing a Kerr medium and two heat baths. Then, we examine the conditions in which the superposition of the coherent states is preserved under the decoherence effects due to the presence of the non-equilibrium environments. For this purpose, we employ the interaction picture to study the decoherence effects. 

Moreover, many approaches have been used to introduce a measure of non-classicality of quantum states \cite{Dodonov}. In this work, we use a measure of non-classicality based on the negativity of the Wigner function which exhibit quantum interference of the system \cite{Kenfack}. It should be noted that the Wigner function can be measured experimentally \cite{Banaszek,Breitenbach}, including the measurements of its negative values \cite{Kurtsiefer,Lvovsky}.

The paper is organized as follows. In section II, we define our model. Then, the formalism for generating of SCSs in MZ interferometer is introduced. Moreover, we calculate the Wigner function for the generated cat states. In section III, we consider two regime to analyze the results of our calculations. Then, we draw various plots in order to better understand the dynamics of SCSs in some physical situations. In section IV, we discuss about physical feasibility of our proposed experiment. Finally, the results are discussed in the concluding remarks section.

\section{Model}

We consider a MZ setup as depicted in Fig.1. The input state $\vert1\rangle $ encounters with two baths at $ T_{H} $ and $ T_{C} $ after its interaction with a non-linear Kerr medium and suffeirng a phase shift $ \theta $. Then, we can write the final state of the system as before detection of the state of the photon in detectors $ \DD_{1,2} $ as
\begin{equation}
\hat{\rho}(t)=\hat{U}^{\dagger}(t)\hat{U}^{\dagger}_{\Kerr}\hat{U}^{\dagger}_{\theta}\hat{U}^{\dagger}_{\BSA}\hat{\rho}(0)\hat{U}_{\BSA}\hat{U}_{\theta}
\hat{U}_{\Kerr}\hat{U}(t)
\end{equation}
with definition $ \hat{U}_{\Kerr} =\exp[i K \tau\hat{a}^{\dagger}\hat{a}\hat{b}^{\dagger}\hat{b}]$, where $ K $ is the Kerr interaction constant related to a third-order non-linear susceptibility $\chi^{(3)}$. Moreover, $(\hat{a},\hat{b})^{\dagger}$ and $ (\hat{a},\hat{b}) $ are the creation and the annhilation operators of the corresponding modes.

\begin{figure}
\centering
\includegraphics[scale=0.38]{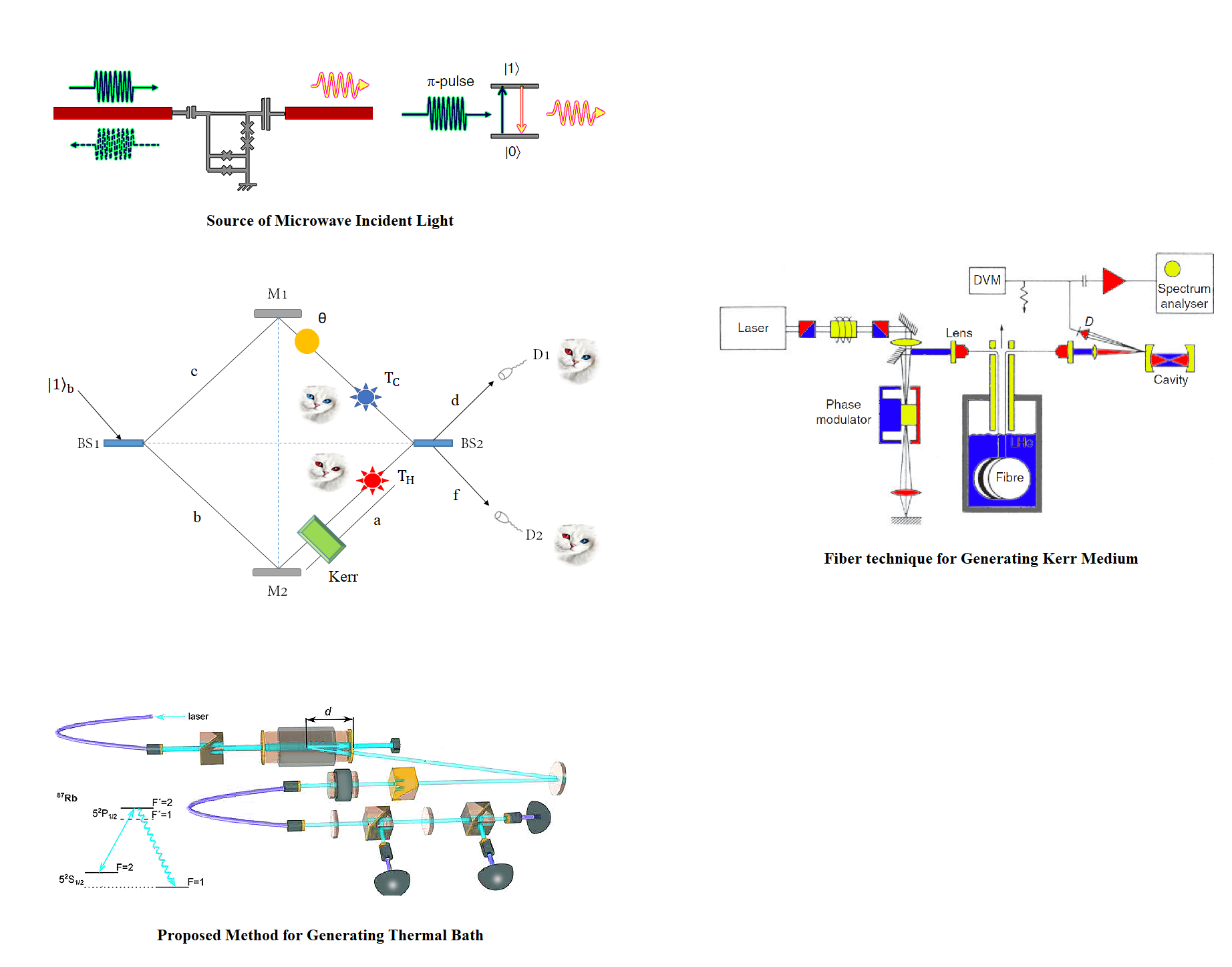}
\caption {Schematic for the proposed MZ interferometer to generate SCSs. Details of the instruments can be found in section IV and references therein.}
\end{figure}

We suppose that the initial state of the non-linear medium is $\vert \alpha\rangle_a$ and the time which the photon passes through the medium is $\tau=l/v$, where $l$ and $v$ are the length of the Kerr medium and the velocity of light in the medium, respectively.  Feeding input mode with $ \vert1\rangle $ in which $K\tau=\pi$, unitary dynamics of the closed system demands that the state of the system before passing through the second beam-splitter $ \BSB $ is
\bq
\vert \psi_0 \rangle=\frac{1}{\sqrt{2}}\big(\vert { -\alpha}\rangle_a\vert 10\rangle_{bc}+i e^{i\theta}\vert \alpha\rangle_a\vert 01\rangle_{bc}\big)
\eq
Then, it is strightforward to show that with $ \theta=\pi $ after $ \BSB $, triggering of the detectors $ \DD_{1,2} $ resluts in superposition of the coherent states $ \vert\psi^{\pm}_{0}\rangle=\dfrac{1}{\sqrt{2}} (\vert\alpha\rangle\pm\vert-\alpha\rangle)$. However, the pitfall of such frameworks in generation of cat states is the delicateness of superpositions against decoherence. Real superposition of classical-like $ \vert \pm\alpha\rangle $ are prone to interaction with their environments and ubiquitous monitoring of the surronding hinders the ability of observing the bizzardness of SCS to be found simultaneously alive and dead. In what follows, we search for an engineered setup of environments that makes it possible to discover the superposition of macroscopic cat states despite the intervention of the environments. \\

\textbf{Formalism}\\

According to Fig.1, we define the total Hamiltonian as $\hat{H}_{\tot}=\hat {H}_0+ \hat{H}_{\Int}$ where
\bq
\hat {H}_0=\hbar \omega_K \hat{a}^\dagger \hat{a} +\hbar \omega \hat{b}^\dagger \hat{b}+\hbar \omega \hat{c}^\dagger \hat{c}+ \sum_i \hbar \omega_{i,H} \hat{d}_{i,H}^\dagger \hat{d}_{i,H}+\sum_i \hbar \omega_{i,C} \hat{d}_{i,C}^\dagger \hat{d}_{i,C}
\eq

\bqali
\hat{H}_{\Int}=\hbar (\hat{a}^\dagger +\hat{a}) \otimes \sum_i c_i(\hat{d}_{i,H}^\dagger  + \hat{d}_{i,H})+\hbar (\hat{b}^\dagger +\hat{b}) \otimes \sum_i c'_i(\hat{d}_{i,H}^\dagger  + \hat{d}_{i,H})+\hbar (\hat{c}^\dagger +\hat{c}) \otimes
\sum_i c''_i(\hat{d}_{i,C}^\dagger  + \hat{d}_{i,C})
\eqali
where $\omega_K$, $\omega$ and $\omega_i$ are the frequencies of the mode which passes through the Kerr medium, the other modes and the $i$-th particle of the bosonic environments, respectively. In addition, $\hat{d}^\dagger $ and $\hat{d}$ are the creation and the annihilation operator of the environmental particles and $c_i,c'_i,c''_i$ are the coupling constants. In Eq. (4), the first term demonstrates the interaction between the non-linear Kerr medium and the heat bath with higher temperature ($T_H$). The second and the third terms denote the interaction between photon and two heat baths.

We use the interaction picture to investigate the evolution of the system in the presence of the non-equilibrium environments. As is well known, time evolution operator $\hat{U}_I(t)$  in the interaction picture can be expanded up to the second order with respect to $\hat{H}_{\Int}$ as
\bqali
\hat{U}_I(t)\simeq 1-\frac{i}{\hbar} \int_0^t \D t_1\hat{H}_{\Int}(t_1)-\frac{1}{\hbar^2}\int_0^t \int_0^{t_1}\D t_1 \D t_2 \hat{H}_{\Int}(t_1)\otimes \hat{H}_{\Int}(t_2)
\eqali
where $\hat{H}_{\Int}(t)$ is the Hamiltonian in the interaction picture. The evolution of the state in the interaction picture can be written as  $ \hat{\rho}_I(t)=\hat {U}_I (t) \hat{\rho}_0 \hat{U}_I^\dagger (t) $, where $\hat{\rho}_0=\vert \psi^{+}_0\rangle \langle \psi^{+}_0\vert \otimes \hat{\rho}_{T_H} \otimes \hat{\rho}_{T_C}$ and $\hat{\rho}_{T_H},\hat{\rho}_{T_C}$ are thermal density matrices of the baths which can be defined as $ \hat{\rho}_{\Th}=\Sigma_{n=0}^{\infty}\dfrac{N^{n}}{(1+N)^{n+1}}\vert n\rangle\langle n\vert $ with $ N=\dfrac{1}{\exp (\hbar\omega / k_{B}T)-1} $. To complete derivation of the reduced dynamics of photon-cat state, we should trace over the environments. Spectral density function, defined as $ J(\omega)=\sum_{0}^{\infty}c_{i}^{2} \delta(\omega_{i}-\omega_{e})$, mediates summerizing properties of the environment in closed mathematical forms where $ c_{i} $ stands for coupling constant between the system and the environment and $ \delta(\omega_{i}) $ is delta function. Moreover, there are different functions for $ J(\omega_{e}) $ that can be written generally as \cite{Taka}
\bq
J(\omega_e)=J_0 \omega_e \big(\frac{\omega_e}{\lambda}\big)^{s-1} \big(1-\frac{\nu_c}{\omega_e}\big)^{\sigma -1}\exp(-\omega_e/\lambda) \theta (\omega_e-\nu_c)
\eq
where $J_0$ is a positive constant which quantifies the strength of the interaction between the environment and the system, $s$ and $\sigma$ are the positive constants which depend on the nature of the environment, $\theta (\omega_e-\nu_c)$ is the step function, $\lambda$ (cut-off frequency) and $\nu_c$ are the  frequencies which maximizes and mimimize $J(\omega_e)$, respectively. In the case of $\nu_c=0$, the spectral density in Eq. (6) can be classified according to the value of $s$ as follows

\bqali
&0<s<1  \hspace{0.9cm}\text{sub-Ohmic environment}\\
&s=1  \hspace{1.5cm}\text{Ohmic environment}\\
&s>1  \hspace{1.5cm}\text{super-Ohmic environment}
\eqali

Trnasforming back to Schr\"{o}dinger picture and applying $ \hat{U}_{\BSB} $, the density matrix of the cat state after detection of the state of photon, can be given by 
\bqali
\hat{\rho}(t)_{\cat}=\Tr_{e}\Tr_{\ph}\big[\hat{U}^{\dagger}_{\BSB}\hat{U}^{\dagger}(t)\hat{U}^{\dagger}_{\Kerr}\hat{U}^{\dagger}_{\theta}\hat{U}^{\dagger}_{\BSA}\hat{\rho}(0)\hat{U}_{\BSA}\hat{U}_{\theta}\hat{U}_{\Kerr}\hat{U}(t)\hat{U}_{\BSB}\big]
\eqali
where $ \Tr_{e}\Tr_{\ph} $ denotes the trace operation on the environments degrees of freedom and the output photon, respectively.

To show that the superposition of alive-dead states has been survived from interaction with surronding, we can survay the behavior of its Wigner function. Negativity of the Wigner function is a trustworthy evidence admits that the macroscopic superposition of SCS is still present despite the probable destruction role of the environment. Our calculations show that the Weyl function of the SCSs, $\Tr(\hat{D}(\gamma)\hat{\rho}(t)_{\cat})$, where $ \hat{D}(\gamma) $ is displacement operator, after triggering of $ \DD_1$  can be written as (see Supplementary Information)

\begin{equation}
\Upsilon(\gamma)=\langle\alpha\vert\hat{F_{1}}(t)\vert\alpha\rangle+\langle\alpha\vert\hat{F_{2}}(t)\vert-\alpha\rangle+\langle-\alpha\vert\hat{F_{3}}(t)\vert\alpha\rangle+\langle-\alpha\vert\hat{F_{4}}(t)\vert-\alpha\rangle+c.c
\end{equation}
where $ c.c $ stands for complex conjugate. Moreover, $ \hat{F}_{i}(t)$s have the following form
\begin{equation}
\hat{F}_{i}(t)=\hat{f_{i}}^{(0)}+\int_{0}^{t}\int_{0}^{t}\D t_{1}\D t^{\prime}_{1}\hat{f_{i}}^{(2)}(\hat{a},\hat{a}^{\dagger},\chi_{T_{C}},\chi_{T_{H}})-\int_{0}^{t}\int_{0}^{t_{1}}\D t_{2}\D t_{1}\hat{f_{i}}^{\prime(2)}(\hat{a},\hat{a}^{\dagger},\chi_{T_{C}},\chi_{T_{H}})
\end{equation}
where $ \hat{f_{i}}^{(0)} $ comes from the unitary part of the evolution. The influence of the environments is infolded in $ \hat{f_{i}}^{(2)} $ and $ \hat{f_{i}}^{\prime(2)} $ terms that represent the interaction between the cat and two baths in different temperatures to the second order. The explicit form of $ \hat{f_{i}}^{(2)} $s can be found in Supplementary Information. Also, $ \chi_{T_{C,H}} $ denotes the correlation functions of the bosonic baths where defined as
\begin{align}
\chi_{T_{C,H}}&=\Sigma_{0}^{\infty}c_{i}^{2}\Big(e^{i\omega_{i}(t-t^{\prime})}\langle \hat{d^{\dagger}}_{i,H(C)}\hat{d}_{i,H(C)}\rangle_{\hat{\rho}_{T_{H(C)}}}+e^{-i\omega_{i}(t-t^{\prime})}\langle \hat{d}_{i,H(C)} \hat{d^{\dagger}}_{i,H(C)}\rangle_{\hat{\rho}_{T_{H(C)}}}\Big)\nonumber\\
&=\int_{0}^{\infty} \D \omega_{e} J(\omega_{e})\Big(\coth(\dfrac{\hbar\omega_{e}}{2k_{B}T_{H(C)}})\cos\omega_{e}(t-t^{\prime})-i\sin\omega_{e}(t-t^{\prime})\Big)
\end{align}

As a consequence, the Wigner function for the generated cat states can be resluted by integral $ \int \D^2 \gamma\exp(\gamma^{\star}\beta-\gamma\beta^{\star}) \Upsilon(\lambda)$ and its general form can be written as (see Supplementary Information)
\bq
W(\beta, \beta^*)=\sum_{i=1}^4 \mu_i+2\re \sum_{i=1}^{18} \Theta_i(t)\mu_i
\eq
where the first term and the second term denote the quantum contribution and the effects of the environment, respectively. The same representation could be established in a similar way for the cat state prepared in $ \DD_{2} $. In the following, we scrutinize the outcomes of analysis based on what can be extracted from the Wigner functions and their integrated form in the momentum space. Results are summed up according to the two limits $ \omega_{K}\simeq \omega_{e} $ and $ \omega_{K}\ll \omega_{e} $ where the correlation functions can be obtained analytically.

\section{Results}

To be more specific, at first, we give the results for the light in microwave band where the related frequencies are not very difficult to achieve experimentally. The frequency of the incident light is set to $ 10^9 $ Hz. Since the spectral density function is broader for high temperature baths, the cut-off frequency has a lower value for the bath in $ T_{H} $. We have set $ \lambda_{C}=2\lambda_{H} $. A cut-off frequency of the order of THz is considered for thermal bath at high temperature. Frequency of Kerr medium has been assumed as the same of the incident light. Temperatures of thermal baths are tuned in $300$ K and $ 100$ K, respectively and oupling constants are assumed $J_{0}\approx 0.1 $. 

The Wigner functions in Fig.2 are summarized the behavior of SCSs when they live in a non-equilibrium environment. Also, we have surveyed the behavior of coherency between the alive-dead states of the cat for the spectral density in Ohmic, sub-Ohmic and super-Ohmic environments, where $s$ in Eq. (7) should be set as $ 1-0.5-2$, respectively. 
 As is clear in Fig.2, there is a negativity area in the Wigner function which exhibit the non-classicality nature of the prepared cat state, in spite of the presence of decoherence effects. Also, it can be inferred that the quantumness of the cat states are robust against decoherence as time is going on. For instance, in the case of $\lambda_{H}t=1000$, if we assume $T_H=300$ K and $T_C=100$ K, the quantum superposition is still alive though the thermal baths has been kept at high temperature limit.

In Fig.3 to Fig.5, $ \int W(x,p) \D p $, which represent the quantum nature of the cat states in the momentum space, are depicted for various physical parameters of the Ohmic, super-Ohmic and sub-Ohmic environments. Oscillations, as the signature of quantum behaviour of the system, confirm that non-equilibrium characteristic of the system-environment interaction makes respiration of the cat between alive or dead states. To facilitate the representation of the mathematical relations, we define dimensionless parameters as $ \varepsilon=\lambda_{H}t $, $ \delta=\omega_K / \lambda_H $ and $ \kappa_H=\hbar \lambda_H / 2k_B T_H $. $ T_{H}>T_{C} $ and $ \lambda_{C}>\lambda_{H} $ demands that $ \kappa_{C}>\kappa_{H}$. Pair values $ \kappa_{C}=0.9, \kappa_{H}=0.9 $ and $ \kappa_{C}=0.9,\kappa_{H}=0.45 $ are used to study the influence of temperature on the dynamics of the system. The figures are depicted for various orders of time parameter $ \varepsilon $. In addition, we consider two regimes for our study. In the first regime, we assume that the system and the environments are in resonance where, $\omega,\omega_{K} \simeq\omega_e $. In the second regime, we assume that the frequency of the environment is much larger than the frequency of the system, $\omega, \omega_K\ll\omega_e $. These regimes  are recruited as reasonable limits of frequencies in which the baths can monitor the state of the cat effectively and the correlation functions of the environment can be analytically solved (see Supplementary Information). The value of the phase is set as $ \theta=\pi $ and $ \theta=\pi / 4$ to study the effect of the change in the phase on the appearance of oscillation patterns. $\delta=0.01 $ is supposed, wherever it was necessary. Since, dimensionless parameters are used, the quantum nature of SCSs can be pursued for different region of frequencies and temperatures. 

Another interesting feature of the Wigner functions is non-symmetric rate of decays for the classical contributions that are in correspondence with two crests in both sides of the negative regions. When one bath is coupled to an open system, it is usually seen in different contexts that the rate of decays in populations is approximately the same. Taking non-equilibrium effects into account can change the play. It can fix the dynamics of the system in regions in which the decay in diagonal elements synchronizes with the robustness of coherences against the demolition role of the environment. Since, now the populations can couple to the environments in a very complicated way (see related terms in Supplementary Information), the mutual synergy between populations and coherences can guide the cat to access to new-found ways for surviving from the destructive role of the environments.

\begin{figure}
\label{2}
\centering
\includegraphics[scale=0.3]{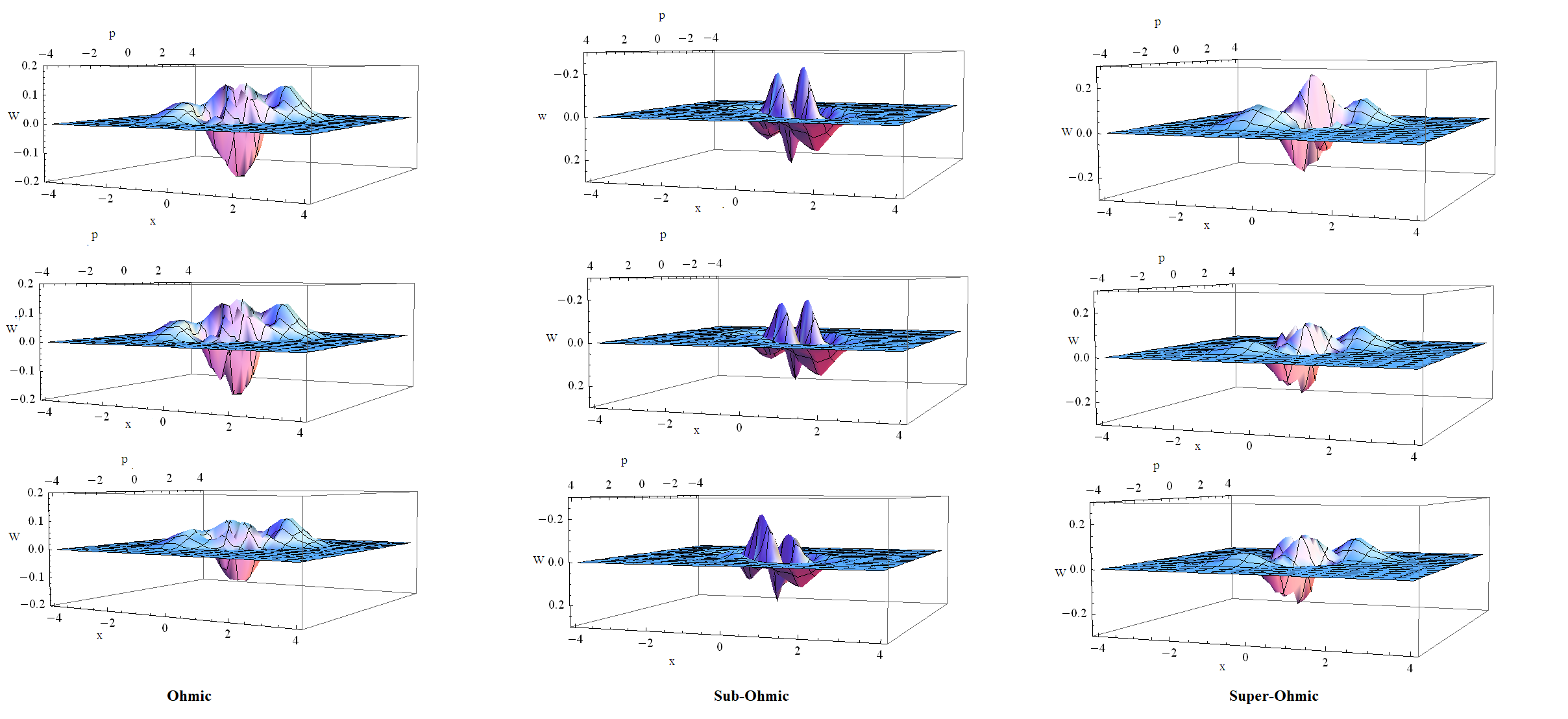}
\caption {The Wigner functions of the cat states are depicted for (Left to Right) Ohmic, super-Ohmic and sub-Ohmic environments. Time goes on from up to down as $\lambda_{H} t=1,10,1000$. Frequency of incident light is in the Microwave band $ \omega=10^{9}$ Hz and supposed to be comparable with the frequency of the Kerr medium. $ J_{0}=0.1 $, $T_{H}=300$ K and $T_{C}=100 $ K are assumed. Cut-off frequencies lie in THz band region.}
\end{figure}

At first sight, it seems that working in two regimes leads to different symmetries in the appearance of oscillations. In resonance limit $ \omega_{K}\simeq \omega_{e} $, oscillations resembles odd functionality with respect to time parameter $ \varepsilon $. In the other limit, the oscillations seems to have even functionality and larger amplitudes. The change in phase $ \theta $ just affect the convexity of the oscillation and has not a major effect on the presence of quantum beats. For $\pi/4$, the oscillation patterns are more regular and the frequency of the cat states, in their open quantum system definition, increases specially in the resonance regime. This means that the quantum nature of the system is much more preserved. Also, an increase in $\kappa_H$, by decreasing in $T_H$ or increasing in $\lambda_H$, leads to a decrease in the height of the summit of the oscillation specially in the resonance regime. It is also concluded that in non-equilibrium framework, $ \omega_{K}\simeq \omega_{e} $ limit and for Ohmic and super-Ohmic environments, the oscillations are more alike to what is seen for isolated quantum systems.  

\begin{figure}
\label{3}
\centering
\includegraphics[scale=0.21]{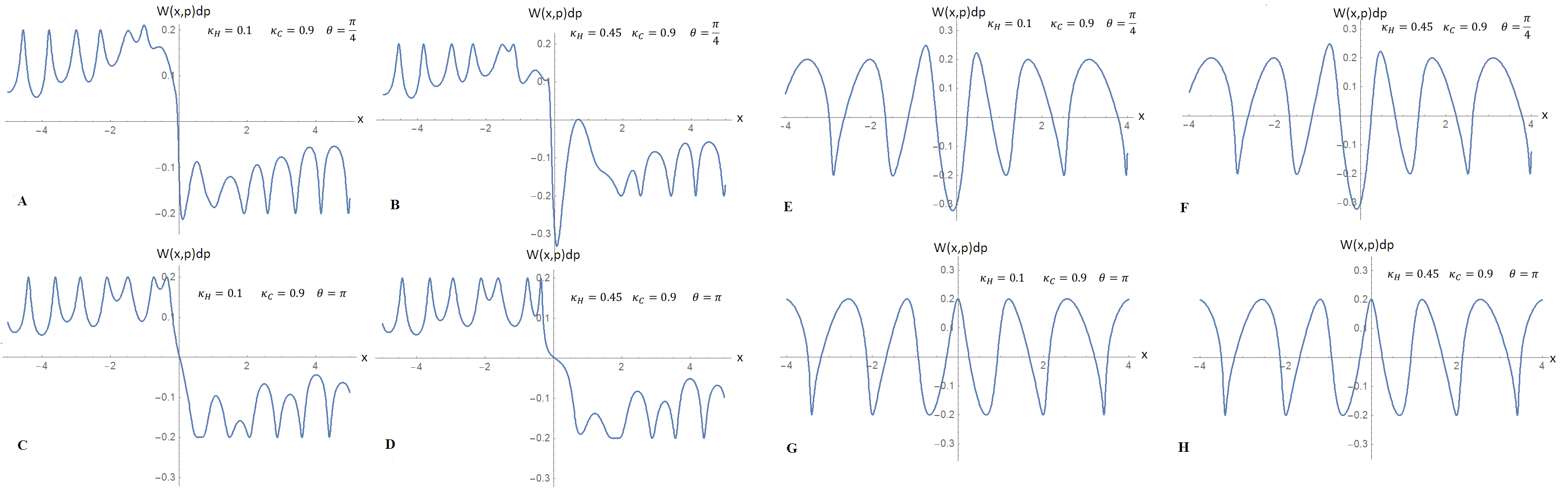}
\caption {Imaginary part of $ \int W(x,p)\D p $ is depicted for an Ohmic environment. In left part
(A to D) $ \omega_e\simeq\omega_{K} $ and in right part (E to H) $ \omega_K\ll\omega_e $ is applied. Values of the other parameters are specified on the plots. All figures are depicted for $ \varepsilon=100 $.}
\end{figure}

\begin{figure}
\label{4}
\centering
\includegraphics[scale=0.21]{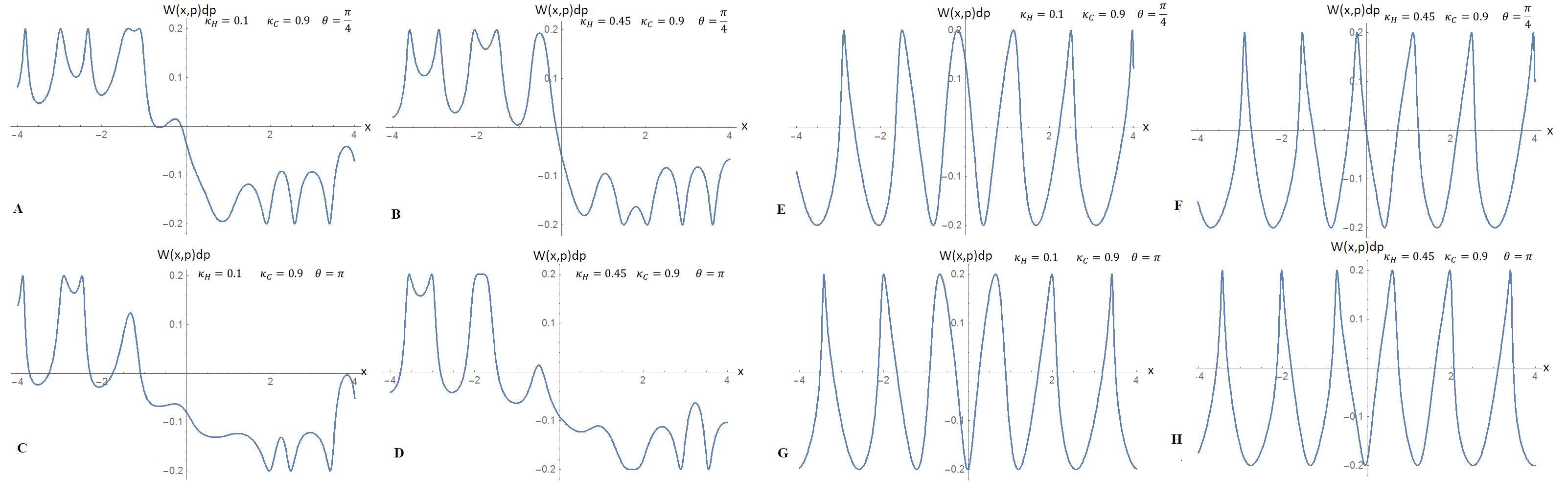}
\caption {Imaginary part of $ \int W(x,p)\D p $ is depicted for a super-Ohmic environment. In left part
(A to D) $ \omega_e\simeq\omega_{K}$ and in right part (E to H) $\omega_K\ll\omega_e$ is applied. Values of the other parameters are specified on the plots. All figures are depicted for $ \varepsilon=100 $.}
\end{figure}

\begin{figure}
\label{5}
\centering
\includegraphics[scale=0.21]{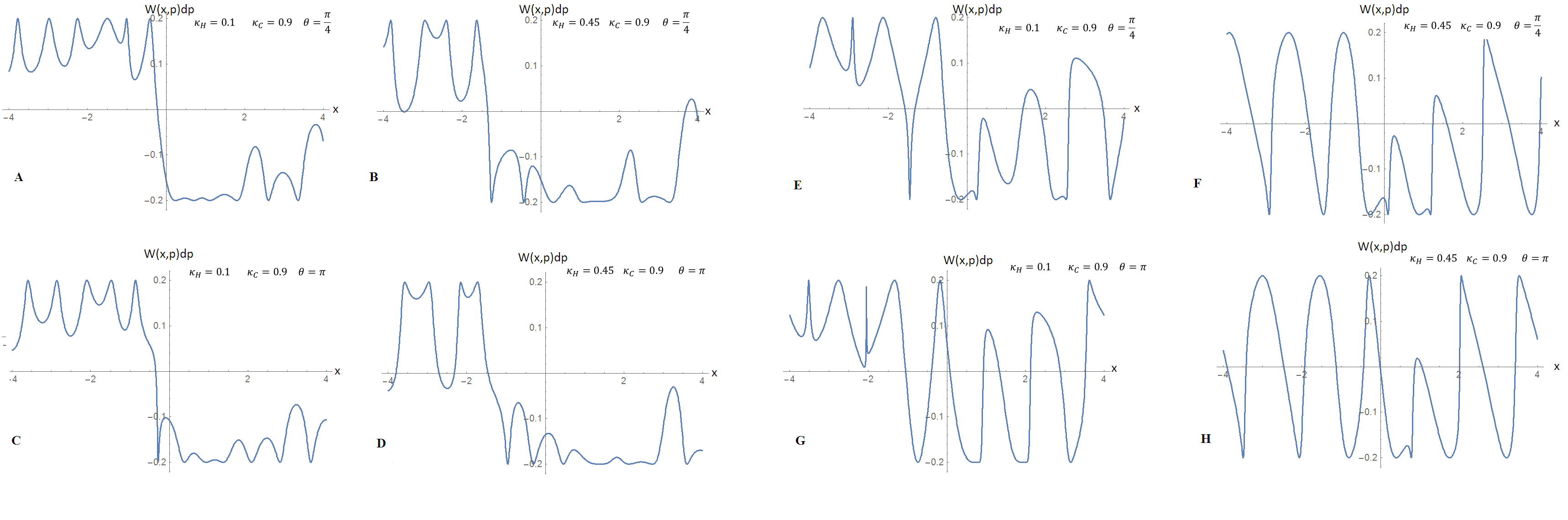}
\caption {Imaginary part of $ \int W(x,p)\D p $ is depicted for a sub-Ohmic environment. In left part
(A to D) $ \omega_e\simeq\omega_K $ and in right part (E to H) $ \omega_K\ll\omega_e$ is applied. Values of the other parameters are specified on the plots. All figures are depicted for $ \varepsilon=100 $.}
\end{figure}

In Fig. 6, one can follow the effect of the exchange in the place of the two baths, where the cold bath is entangled with the initial state of the cat for Ohmic, super-Ohmic and sub-Ohmic environments. As is clear, this exchange makes the oscillation patterns more symmetrical, specially in an Ohmic environment. However, the general form of the oscillations does not change significantly from the previous one. 

\begin{figure}[H]
\label{6}
\centering
\includegraphics[scale=0.28]{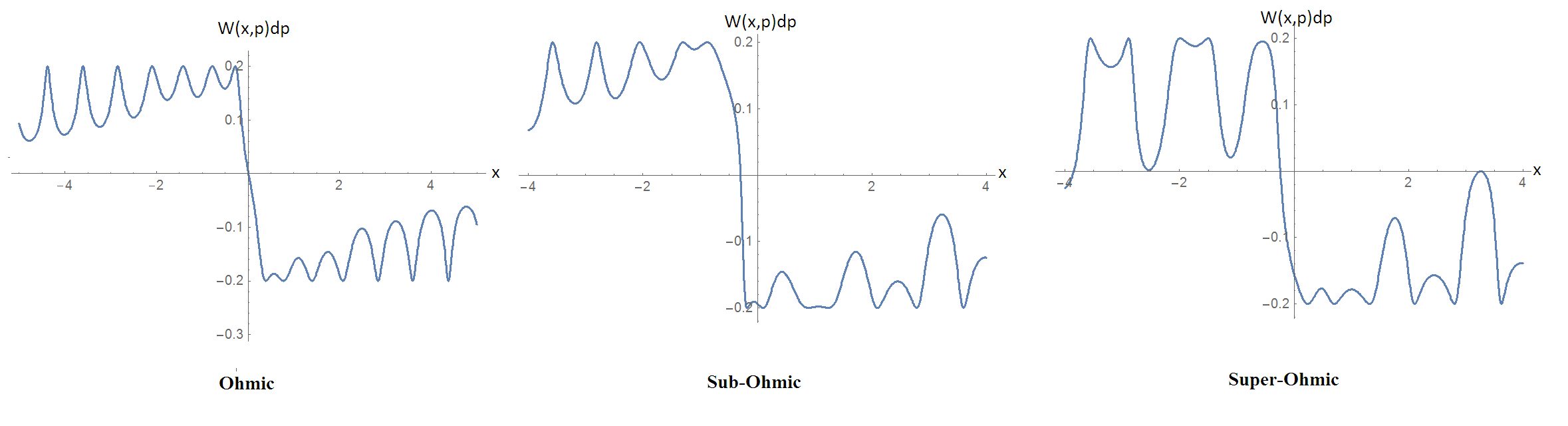}
\caption {Imaginary part of $ \int W(x,p)\D p $ is depicted for Ohmic, sub-Ohmic and super-Ohmic environments from left to right where the cold bath is entangled with the initial state of the cat. Parameters are set as $ \kappa_{H}=0.45, \kappa_{C}=0.9, \theta=\pi,\varepsilon=100 $.}
\end{figure}

\section{General Conditions for Physical Feasibility of Non-equilibrium Respiration}

To have a realistic picture of the problem, in this section we survey the possibility of physical implementation of the model. MZ is a building block that has paramount importance in tests of quantum optics. There are various ways of setting up MZ interferometers \cite{M1}. The main focus of our work is to engineer the destructive role of the environment in a way that quantum features of the system could be revived. It is not surprising that at sufficiently low temperatures the quantum behavior of the system could be preserved. Observing the quantumness at almost high temperatures may have critical importance for experts in area of the quantum optics and quantum information. To have physically meaningful effects of the thermal baths at sufficiently high temperature limits, the frequency of incident light may be adjusted in the microwave band i.e. 1 GHz to over 100 GHz. Z. Peng et al. proposed the operation of a tuneable on-demand microwave photon source based on a fully controllable superconducting artificial atom strongly coupled to an open-ended transmission line \cite{M2}. Superconducting quantum systems provide a novel basis for the realization of microwave photon sources \cite{M3,M4,M5,M6,M7}. Spontaneous non-degenerate parametric down conversion also can be hired to generate a single photon state at the input mode of the MZ interferometer \cite{M8}.  

One of the most critical challenges for the realization of the model is providing a significant non-linearity to obtain a large separation between two components of the generated SCS. Maximum separation for the case in which $ K\tau=\pi $, requires a large Kerr non-linearity or a long non-linear medium such as a long optical fiber. It has been shown that to generate a phase shift on the order of $ \pi $ for an optical frequency of about $\omega=5\times10^{14}$ rad/sec, an optical fiber with length of 3000 km should be used \cite{K1}. However, generated state after outgoing the non-linear medium will be completely decohered because of the significant losses of coherence in the baths. To overcome this difficulty, a large amplitude of the initial coherent state can be used with a short interaction time \cite{K2}.  If the amplitude of the coherent state is large, the same amount of separation can be obtained in the phase space even though $ K\tau $ is much smaller than $ \pi $. Using fiber-optic Kerr squeezing can impressively reduce the quantum noise and improve the amplification of Kerr effect \cite{K3,K4}. Rotation of the polarization of emitted light resulted from atomic vapor can reinforce non-linearity in the initial isotropic $ \chi^{(3)} $ medium by amplifying squeezing up to $-0.85$ dB in the experiment \cite{K5}. 

A thermal source obtains by collecting emitted light of ionization processes in hot vapor of atoms like mercury and hydrogen \cite{f1}. Passing laser beams through a fast rotating grounded glass is another useful way of making thermal source in MHz  bandwidth \cite{f2,f3}. The scattered incident light disperses in all directions randomly and its phase and amplitude will be random variables accordingly. Distribution of the scattered light obeys from a quasi thermal density operator. Recently, Mika and coworkers have suggested a generic method for constructing ideal thermal light by excitation of a warm vapor of Rubidium atoms by a laser beam with 795 nm wavelength \cite{f4}. The anti-Stokes field generated by  $ 5^{2}S_{1/2}\rightarrow 5^{2}P_{1/2} $ Raman transition excited by the laser beam becomes polarized in adjusted angles to suppress the detection of the light coming from the residual laser reflections. The proposed method extends the bandwitdth of thermal light frequency to GHz regions. Laser detuning techniques and nanomechanical heat engines are contemporary methods for regulating the temperature of heat baths in desired limits \cite{f5,f6}.

\section{Concluding Remarks}

The non-appearance of quantum features in macroscopic world, is one of the most fundamental challenges in the heart of quantum physics. It was first addressed by the well-known Schr\"{o}dinger's Cat paradox. Standard quantum mechanics flees from responding to the question of finding cat both alive-dead states simultaneously, by conceal it behind the collapse postulate or proposing the inevitable role of the observer in doing measurements. There was everlasting efforts to overcome this dilemma  by presenting physical interactions that result to macroscopic realism from microscopic context. Decoherence suggests that ubiquitous interaction of the real state of the cat with its environments hinders observation of macroscopic superposition. Then, it is reasonable to ask what is the boundary on which respiration of cat is going to be stopped due to such interaction.

In this effort, we examined the possibility of making superposition of alive-dead states when the environments far from equilibrium are hired in the arms of the Mach-Zehender interferometer. In our proposed setup, a non-linear Kerr medium is the source of generation of the cat states. As can be deduced from our results, when the interaction of the open system is studied in non-equilibrium conditions, it could be possible to revive the quantum beats, even at high temperatures limits and in the presence of decoherence effects. As a consequence, it seems that destructive role of the environments on the coherency of the system, can be engineered in a way that in which quantum behaviour of the system still has chance to be revived. We have investigated the types of different environments and their physical parameters for finding optimized situations in which quantum oscillations can be clearly observed. 

SCSs play a prominent role in doing fundamental tests of quantum mechanics and also to probe the boundaries on them the quantum to classical transition takes place. At the other side, preserving superposition in experiment is critical for making physical realization of quantum computers. Then, probably, our work may have useful consequences in extending the realm in which Schr\"{o}dinger's Cat could be able to exist. If so, it equipped us with new insights to resolve some fundamental and practical questions in nowadays theory of open quantum systems.

\end{document}